\documentclass[twocolumn,superscriptaddress,longbibliography,aps,preprintnumbers]{revtex4-2}

\usepackage{graphicx,color}
\usepackage{xcolor}
\usepackage{bm,bbm,amsmath,amssymb,braket}
\usepackage{epstopdf}
\usepackage{relsize}
\usepackage{epstopdf}
\usepackage{comment}
\usepackage{soul}
\usepackage{float}
\usepackage{lipsum}
\usepackage[urlcolor=blue,colorlinks=true,citecolor=blue,
	linkcolor=red,pdfstartview={FitH},bookmarks=false]{hyperref}
\usepackage{MnSymbol}
\usepackage{stmaryrd}
\usepackage{mathtools}

\graphicspath{{Figures/}{./Figures/}{.}{./paper_figures/}}
\setstcolor{red}

\definecolor{purple}{rgb}{0.5, 0., 0.8}

\definecolor{persianblue}{rgb}{0.11, 0.22, 0.73}

\definecolor{darkred}{RGB}{182, 24, 24}

\begin{document}

\title{Multicopy teleportation and quantum program retrieval on superconducting quantum hardware}

\author{Yousef Mafi}\email{Email: yousef.mafi@tuni.fi}
\affiliation{Computational Physics Laboratory, Physics Unit, Faculty of Engineering and Natural Sciences, Tampere University, P.O. Box 692, FI-33014 Tampere, Finland}

\author{Ali G.  Moghaddam}\email{Email: ali.moghaddam@aalto.fi}
\affiliation{Department of Applied Physics, Aalto University, 02150 Espoo, Finland}
\affiliation{Computational Physics Laboratory, Physics Unit, Faculty of Engineering and Natural Sciences, Tampere University, P.O. Box 692, FI-33014 Tampere, Finland}
\affiliation{Helsinki Institute of Physics P.O. Box 64, FI-00014, Finland}

\date{\today}


\begin{abstract}
    Correction-free teleportation protocols offer a probabilistic route around the limits imposed by the quantum no-programming theorem, but have remained largely theoretical. Here, we experimentally implement and benchmark correction-free multicopy teleportation on noisy superconducting quantum hardware. Key to this is an efficient circuit implementation: by suitably decomposing the unitary that maps the multicopy measurement basis onto the computational basis, we substantially reduce circuit depth and gate count. For one- and two-qubit input states, using $k=2$ and $3$ copies, we observe success probabilities substantially exceeding the single-copy bound with fidelities above $0.85$. We further explore applications to probabilistic programmable quantum processors, in particular the storage and retrieval of unitary operations encoded in quantum program states. Our results establish the practical feasibility of multicopy teleportation and highlight its potential for near-term quantum processors and distributed quantum computing.
\end{abstract}

\maketitle

\emph{\color{blue} Introduction.---}Quantum teleportation is a fundamental primitive of quantum information processing, transferring an unknown quantum state using shared entanglement and classical communication~\cite{Bennett1993teleportation,Bouwmeester1997teleportation,Furusawa1998teleportation,Pirandola2015advances}. Together with closely related protocols such as entanglement swapping and gate teleportation, and complemented by Bell nonlocality as an independent resource, it has become a key building block of quantum technologies, ranging from quantum networks and cryptography to fault-tolerant quantum computation~\cite{Zukowski1993swapping,Gottesman1999gate,Brunner2014nonlocality,Briegel1998repeater,Kimble2008internet,Raussendorf2001oneway,Huhtanen2026swapping,Mafi2026high-dimensional}. In its standard form, teleportation requires a receiver-side correction conditioned on the sender's Bell-measurement outcome. This correction becomes problematic when teleportation is used to implement an unknown quantum operation encoded in a program state. For instance, a unitary $U$ can be stored in the entangled resource state $(\openone\otimes U)\ket{\Phi^+}$ and retrieved through gate teleportation \cite{Gottesman1999gate}. However, the required correction generally depends on the unknown $U$, which prevents a deterministic implementation. This limitation is closely related to the quantum no-programming theorem, which prohibits exact, deterministic universal programming of arbitrary unitaries with finite-dimensional program states~\cite{Nielsen1997programming}. Alternatively, postselecting a single Bell-measurement outcome enables exact correction-free teleportation with success probability $1/d^2$ in dimension $d$, providing a probabilistic way to store and retrieve unknown quantum operations~\cite{Sedlak2019storage}.

A prominent correction-free approach is port-based teleportation (PBT) where, instead of corrections, Bob simply keeps the specific port indicated by Alice’s measurement outcomes and discards all the other ports 
\cite{Ishizaka2008pbt,Ishizaka2009pbt}. But the method comes at the cost of additional entanglement and complex collective measurements. Extensive theoretical developments have established optimal resource states and success probabilities for deterministic and probabilistic variants \cite{Studzinski2017pbt,Mozrzymas2018optimal,Christandl2021asymptotic,Studzinski2022multiport}. More recently, Schur--Weyl duality and its mixed or twisted generalizations have enabled efficient implementations of PBT measurements \cite{Fei2023efficient,Grinko2023efficient,Wills2024efficient}. Despite these advances, experimental implementation on noisy quantum processors remains challenging due to the circuit depth, ancillary resources, and nonlocal operations required for the highly involved collective measurements.

\begin{figure}
    \centering
    \includegraphics[width=0.8\linewidth]{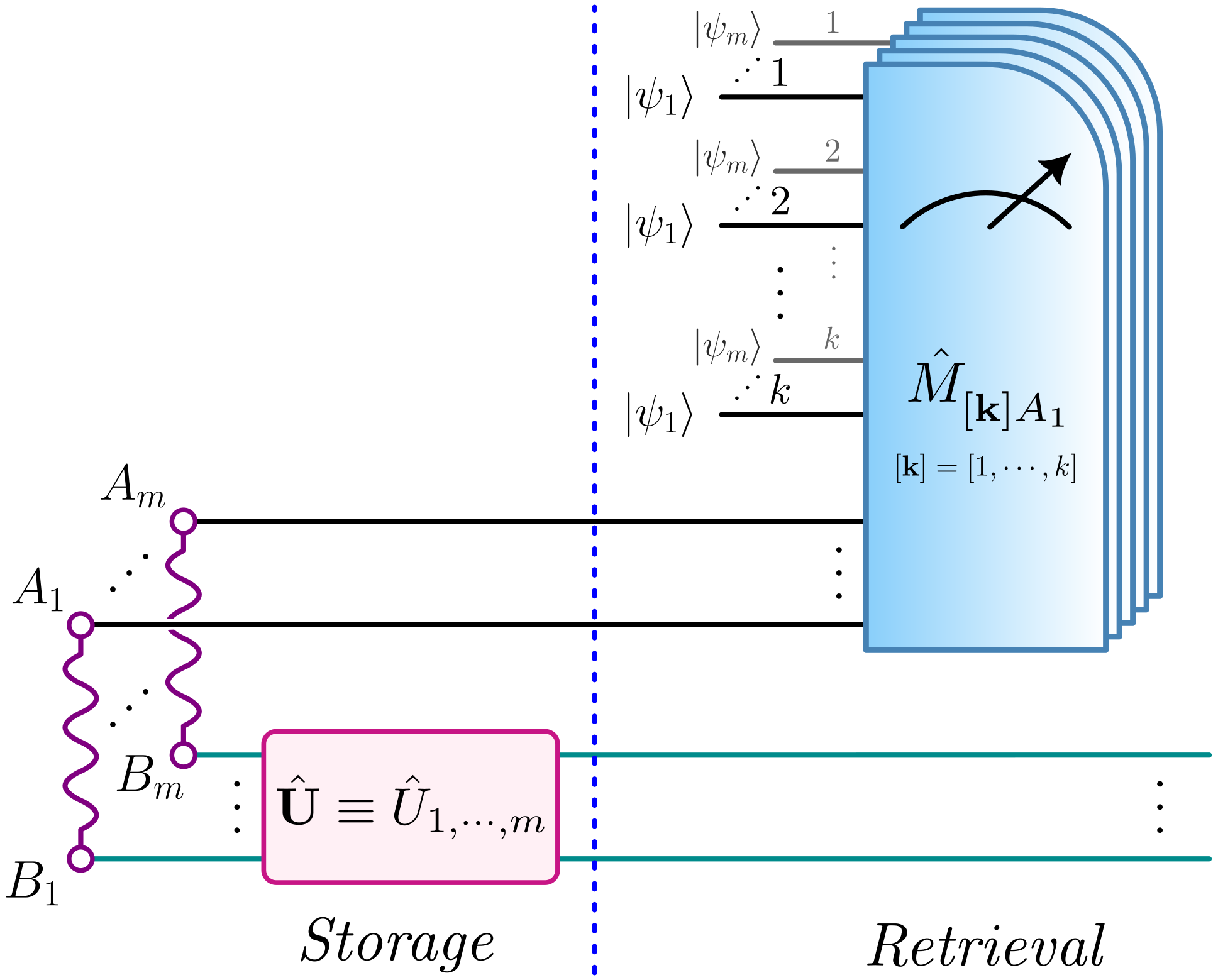}
    \caption{
    \textbf{Generalized multicopy teleportation scheme.}
    During storage, an unknown operation $\hat U=\hat U_{1,\dots,m}$ acts once on Bob's side of
    $m$ Bell pairs $A_j B_j$. During retrieval, a collective measurement
    $\hat M_{[\mathbf{k}]A_{1\dots m}}$ on $k$ copies of the input and Alice's channel qubits,
    with individual probability $p(d,k)$, the exact action of $\hat U$ on the input without
    any correction at Bob's side. }
    \label{fig:1}
\end{figure}

A complementary strategy, known as multicopy teleportation, has recently emerged, in which the additional resource is not shared entanglement but multiple copies of the input state \cite{Grosshans2026multicopyquantum}. Given $k$ identical copies of an arbitrary unknown pure state $|\psi\rangle$, a collective measurement on these copies and the sender's half of a single maximally entangled pair enables exact correction-free teleportation with optimal success probability $p(d,k)=k/[d(k-1+d)]$, where $d$ is the dimension of the teleported system. This probability approaches $1/d$ as $k\rightarrow\infty$, improving upon the conventional $1/d^2$ success rate without requiring additional shared entanglement. The protocol also offers a route to enhancing the probabilistic storage and retrieval of quantum programs encoding unknown operations. A generalized multicopy teleportation scheme and its application to quantum program retrieval are illustrated schematically in Fig.~\ref{fig:1}.

In this Letter, we experimentally investigate multicopy teleportation and its application to quantum program retrieval on superconducting quantum hardware. Using unitary completion of the collective measurement operator followed by quantum Shannon or isometry decomposition of the resulting multiqubit unitary, we construct explicit circuit implementations of the protocol. We observe success probabilities above the conventional correction-free teleportation rate, although hardware noise suppresses this advantage at the largest copy number tested. We also demonstrate the retrieval of stored quantum programs encoding single- and two-qubit operations, with enhanced success probabilities and high fidelities. These results highlight the potential and current hardware limitations of multicopy teleportation for programmable quantum processors and distributed quantum computing.

\emph{\color{blue}Multicopy teleportation.---}The core idea 
in this protocol is to use multiple
copies of the unknown qudit state $\ket\Psi$ by performing collective measurements on the copies and Alice's side of the qudit Bell pair $\ket{\Phi_d^+}:=\frac{1}{\sqrt{d}}\sum_{i=0}^{d-1}\ket{ii}$. By exploiting the permutation symmetry of the copies, the protocol achieves the optimal success probability $p(d,k)$. In this protocol a two-outcome quantum measurement is performed, described by a  positive operator-valued measure (POVM)  
$\{\hat{M}, {\mathbbm 1} - \hat{M}\}$. 
The corresponding collective
measurement operator
is given by \cite{Grosshans2026multicopyquantum},
\begin{align}\label{eq:measurement}
   \hat{M}_{[k]A} = \sum_{i=0}^{R-1} \ket{r_i}\bra{r_i},\qquad R ={k-2+d \choose k-1}
\end{align}
with $[k]=\{1,\dots,k\}$ acting on the $k$ copies
$\ket{\Psi}^{\otimes k}$
and Alice's half of the Bell pair. The vectors 
$\ket{r_i}\in(\mathbb C^d)^{\otimes(k+1)}$
are constructed from the 
orthonormal basis of the symmetric subspace (such as Dicke states for $d=2$) of $k-1$ qudits and a Bell state between one of the input qudits and Alice's half of the Bell pair. Explicitly, they are defined as 
\begin{align}
    \ket{r_i}:=\frac{1}{\gamma_{k,d}} \sum_{a=1}^k 
    \ket{D^{(k-1)}_i}_{ [k] \setminus a }\otimes\ket{\Phi_d^+}_{a A},
\end{align}
where $\gamma_{k,d}=\sqrt{k(k-1+d)/d}$ is the normalization factor, and $\ket{D^{(k-1)}_i}_{ [k] \setminus a }$ denotes the basis state $i$ for the symmetric subspace acting on all $k$ copies except the $a$\textsuperscript{th} one. 
The choice of POVM element $\hat{M}$ 
given by Eq. \eqref{eq:measurement}
maximizes the success probability among all measurements that leave Bob exactly in $\ket{\Psi}_B$ without correction. 

\emph{\color{blue}Quantum circuit implementation.---}As illustrated in Fig.~\ref{fig:1}, the protocol extends naturally to multiple copies of $m$-qubit states, $\ket{\Psi}=\ket{\psi_1}\otimes\cdots\otimes\ket{\psi_m}$. For product states, the measurement can be performed independently on the $j$-th qubits of all copies, which reduces the cost of the protocol. For general (entangled) $m$-qubit states, the measurement must instead act collectively on all qubits of all copies, with $m$ Bell pairs as the entanglement resource. To obtain an explicit circuit, we use unitary completion: we construct a unitary that maps the support of the measurement operator $\hat M$ and its orthogonal complement onto distinct sets of computational basis states.

We first collect the states $\{\ket{r_i}\}_{i=0}^{R-1}$ into the matrix
$\mathcal M=\begin{bmatrix}\ket{r_0} & \cdots & \ket{r_{R-1}}\end{bmatrix}\in\mathbb C^{N\times R}$, with $N=d^{\,k+1}$. Completing them with an orthonormal basis $\{\ket{r'_j}\}_{j=R}^{N-1}$ of the orthogonal complement yields the unitary
\begin{align}
\mathcal W=\sum_{i=0}^{R-1}\ket{i}\bra{r_i}+\sum_{j=R}^{N-1}\ket{j}\bra{r'_j},
\end{align}
which maps each $\ket{r_i}$ to the computational basis state $\ket{i}$ and the complement to the remaining basis states. Applying $\mathcal W$, measuring in the computational basis, and grouping the outcomes $0,\ldots,R-1$ therefore realizes the measurement of $\hat M=\mathcal M\mathcal M^\dagger$. In practice, we obtain $\mathcal W$ from the full QR decomposition $\mathcal M=\mathcal Q\mathcal T$, with unitary $\mathcal Q\in\mathbb C^{N\times N}$. Since $\mathcal M$ has orthonormal columns, $\mathcal T$ is diagonal up to phases, so after absorbing these phases the first $R$ columns of $\mathcal Q$ coincide with $\mathcal M$ and the remaining columns span its orthogonal complement. Hence $\mathcal W=\mathcal Q^\dagger$. Finally, we decompose $\mathcal W$ into single- and two-qubit gates using the quantum Shannon decomposition (QSD) and isometry synthesis~\cite{shende2006,iten2016}. These decompositions, their circuit complexity, and the details of the experimental implementation are discussed after the results.

\begin{figure}
    \centering
    \includegraphics[width=\linewidth]{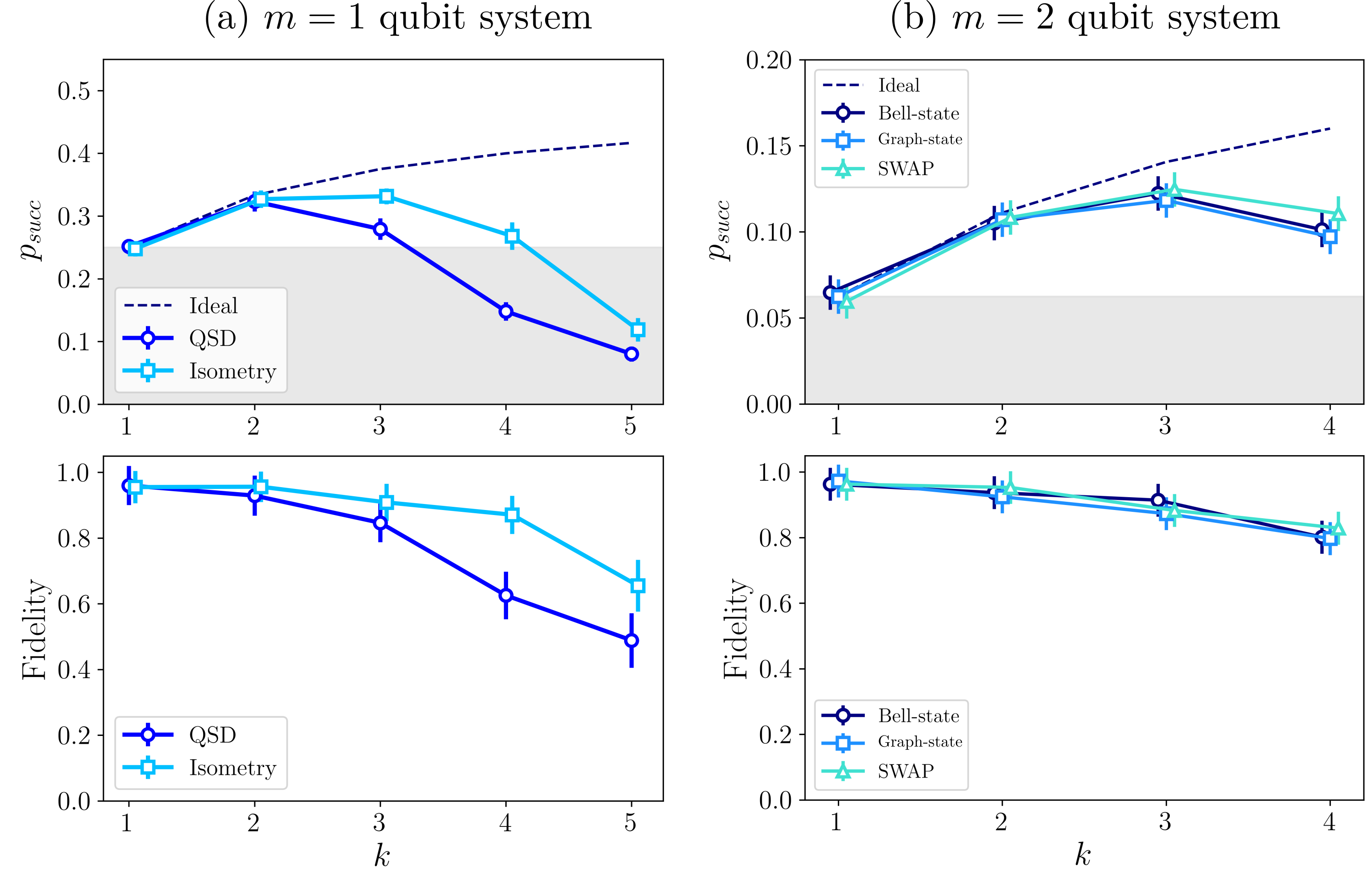}
    \caption{\textbf{Multicopy teleportation on \textit{ibm\_kingston}.} Success probability (top) and fidelity (bottom) as functions of the copy number $k$ for (\textbf{a}) single-qubit teleportation using QSD and isometry decompositions, and (\textbf{b}) two-qubit program retrieval, with Bell-state, graph-state, and SWAP programs encoded in Bob's qubits and applied to the input states $\ket{00}$, $\ket{00}$, and $\ket{01}$, respectively. Dashed lines indicate the ideal success probabilities, while shaded regions mark values below the corresponding single-copy success rates of 1/4 and 1/16. Error bars represent the standard deviation estimated from 20{,}000 shots.}
    \label{fig:2}
\end{figure}

\emph{\color{blue}Multicopy teleportation on real hardware.---}Here, we present experimental results for single- and two-qubit input states $\ket{\Psi}$, implementing $k$-copy teleportation with $k\leq 5$ on \texttt{ibm\_kingston}. Figure~\ref{fig:2} shows the success probability and the fidelity of Bob's final state. Single-copy correction-free teleportation ($k=1$) succeeds with probability $p_{\rm succ}=1/4$ and $1/16$ for $m=1$ and $2$, respectively. Using multiple copies, we observe significantly higher success probabilities for $k=2,3$ with single-qubit states and for $2\leq k\leq 4$ with two-qubit states. The fidelities remain above $90\%$ for $k\leq 2$ and above $80\%$ up to $k=3$ for $m=1$ and $k=4$ for $m=2$. In theory, the success probability keeps increasing with $k$ and asymptotically approaches $1/2$ and $1/4$ for $m=1$ and $2$. In practice, however, the unitary that maps the collective measurement basis, built from symmetric subspaces, onto the computational basis becomes more complex with $k$. The resulting deeper circuits accumulate more noise, so that for $k=5$ both the success probability and the fidelity degrade significantly.

\emph{\color{blue}Quantum program storage and retrieval.---}Next, we use the protocol as a storage-and-retrieval processor for a unitary operation $\hat{\bf U}$ as illustrated in Fig. \ref{fig:1}. In the storage phase, $\hat{\bf U}$ is applied to Bob's half of the shared Bell pair, encoding the operation in the resulting quantum state, which serves as the stored program. In the retrieval phase, a successful measurement $\hat{M}$ on Alice's subsystem and the input copies produces $\hat{\bf U}\ket{\Psi}_B$. Thus, the operation is encoded before the unknown input state is supplied, and its action on that state is retrieved later through the teleportation protocol.

\begin{figure}[!]
    \centering
\includegraphics[width=0.99\linewidth]{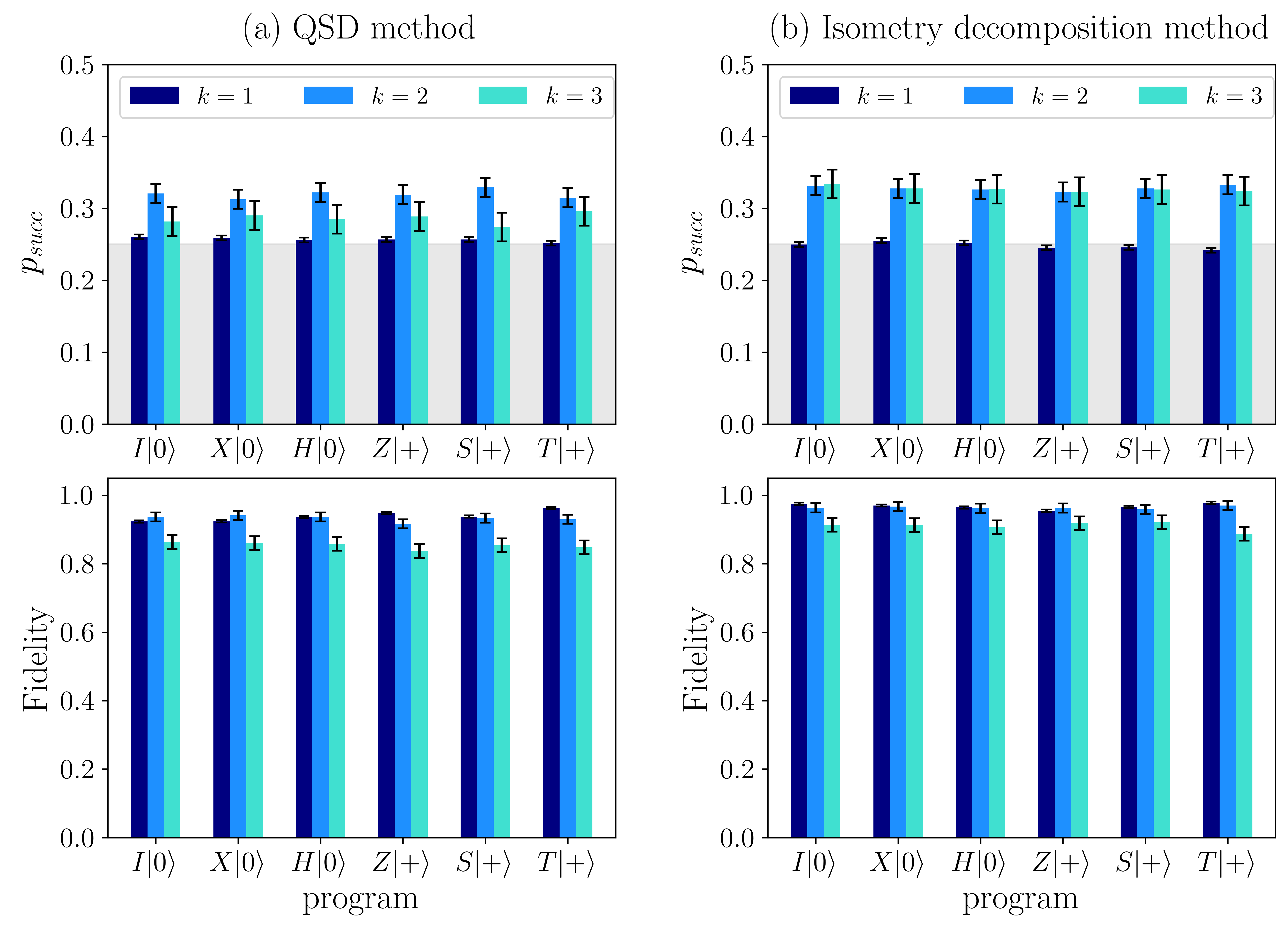}
    \caption{
    \textbf{Quantum program retrieval in a multicopy protocol.}
    Success probability (top) and fidelity (bottom) for six stored program--input
    combinations, retrieved with $k=1,2,3$ copies using (\textbf{a}) QSD and (\textbf{b})
    isometry decomposition.}
    \label{fig:3}
\end{figure}

We store six different single-qubit programs in separate experiments: the identity, Hadamard $H$, Pauli $X$ and $Z$, phase $S$, and non-Clifford $T$ gates. Each program is applied to one input state, as listed in Fig.~\ref{fig:3}, for $k=1,2,3$, and the output is compared with the ideal target state. The success probabilities and fidelities for both decompositions are shown in Fig.~\ref{fig:3}. As in multicopy teleportation, using additional input copies improves the success probability, while the fidelity decreases slightly because the more complex circuits incur greater hardware noise. Isometry decomposition again yields better performance than QSD, consistent with its shallower circuits and consequently reduced exposure to noise.

\begin{figure}
    \centering
    \includegraphics[width=\linewidth]{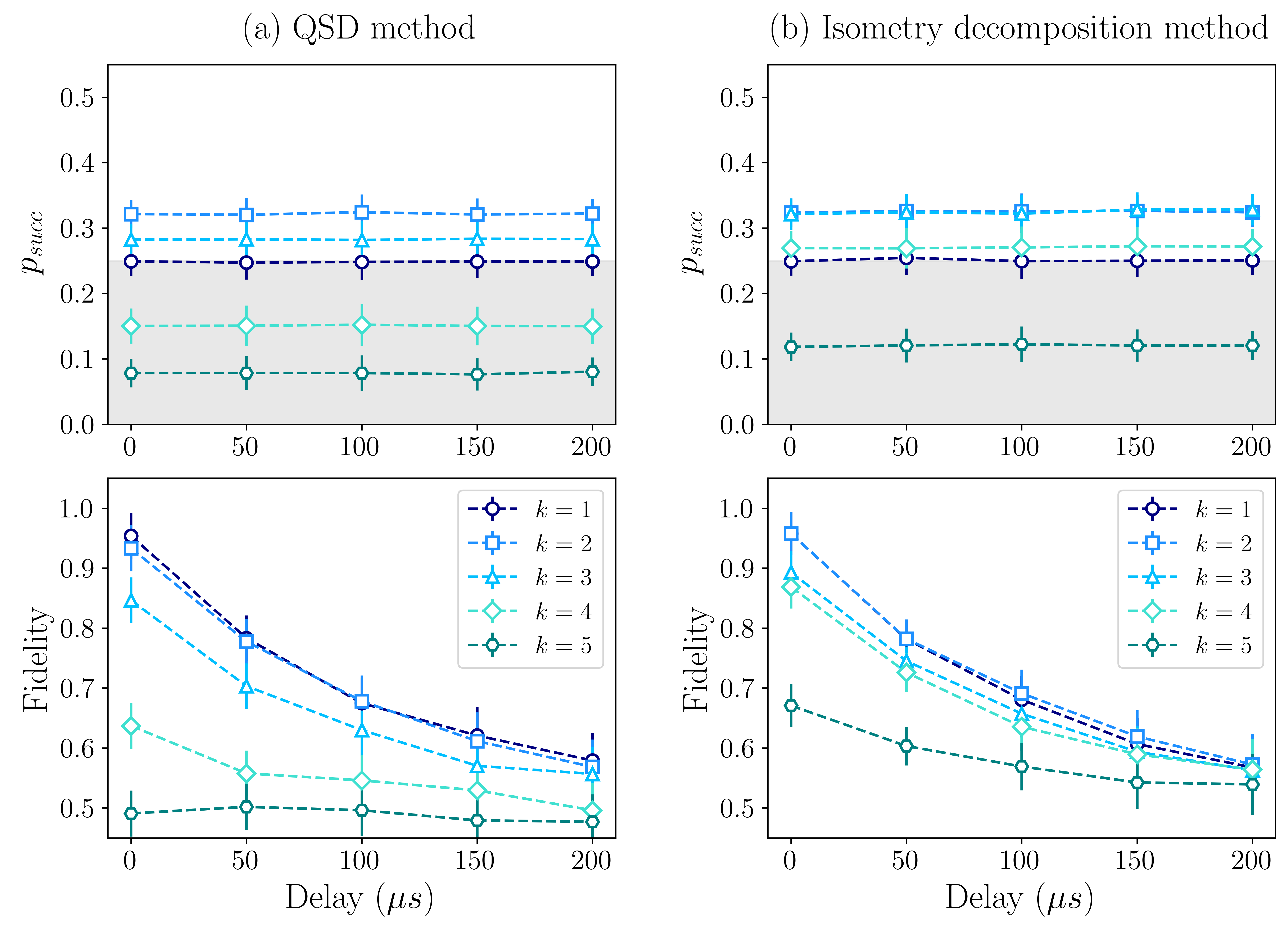}
    \caption{
    \textbf{Program-memory lifetime on \textit{ibm\_kingston}.}
    Success probability (top) and retrieval fidelity (bottom) versus the delay between the
    storage and retrieval phases, for $k=1$--$5$ with (\textbf{a}) QSD and (\textbf{b})
    isometry decomposition. The success rate is essentially delay independent, whereas the
    fidelity decays with the coherence of the stored program state.}
    \label{fig:4}
\end{figure}

In a practical storage-and-retrieval setting, the program state may not be consumed immediately. The waiting time in memory between the storage and retrieval phases leads to decoherence and degradation of the stored program. This is illustrated in Fig. \ref{fig:4} by introducing a delay between preparing the Bell state and performing the measurement on the copies and Alice's subsystem. We observe no change in the success probability, while the fidelity decreases exponentially with the delay $\tau$. This effect can be modeled by assuming a noisy Bell pair,
\begin{equation}\label{eq:noisy-bell}
    \rho_{AB}^{(\lambda)} = (1-\lambda)\ket{\Phi_d^{+}}\bra{\Phi_d^{+}}_{AB}
    +\lambda\,\frac{\mathbbm{1}_{AB}}{d^2},
\end{equation}
which corresponds to either subsystem undergoing a depolarizing channel ${\cal D}_\lambda(\rho)=(1-\lambda)\rho+\lambda\;\mathbbm{1}/d$.
Assume that Bob's subsystem is affected by this noise before the measurement. Since the measurement operator ${\hat M}$ acts only on Alice's subsystem and the $k$ copies, its action commutes with the noise channel on Bob's subsystem. Therefore, the success probability remains unchanged, while the output fidelity decreases as if Bob's subsystem had undergone depolarization after the measurement:
${\cal F}_\lambda=1-(d-1)\lambda/d$,
corresponding to
$\rho_B^{\rm out}=(1-\lambda)\ket{\psi}_B\bra{\psi}+\lambda\,\mathbbm{1}/d$.
The growth of noise during storage can be described phenomenologically by
$\lambda(\tau)=1-e^{-\tau/\tau_c}$, with $\tau_c\approx(1/T_2^{A}+1/T_2^{B})^{-1} \approx 65~\mu{\rm s} $,
where $T_2^{A}$ and $T_2^{B}$ denote the coherence times of the two memory qubits; the lifetime of the stored program is therefore determined by the dephasing of the shared Bell pair. The fidelity then approaches $1/d$ exponentially on the characteristic time scale $\tau_c$ in agreement with the results shown in Fig.~\ref{fig:4}.

\emph{\color{blue}Circuit synthesis and complexity.---}We now return to the two methods for synthesizing the measurement unitary $\mathcal W$. The first treats $\mathcal W$ as a generic $n$-qubit unitary, with $n=k+1$, and decomposes it using the QSD. The second exploits the structure of the measurement. Only the first $R$ columns of $\mathcal W^\dagger$, the vectors $\ket{r_i}$, determine the measurement, and any completion of their orthogonal complement implements the same POVM $\{\hat M,\openone-\hat M\}$. It therefore suffices to synthesize an isometry from $q=\lceil\log_2 R\rceil$ to $n$ qubits.

The two approaches differ in their scaling. Any decomposition of an arbitrary $n$-qubit unitary requires at least $\lceil(4^{n}-3n-1)/4\rceil$ CNOT gates. The QSD uses $\tfrac{23}{48}\,4^{n}-\tfrac{3}{2}\,2^{n}+\tfrac{4}{3}$ CNOT gates, which is within a factor of about two of this bound and gives $N_{\rm QSD}=\Theta(4^{k})$ \cite{shende2006}. The isometry decomposition instead requires approximately $2^{\,n+q}=2^{\,k+1+\lceil\log_2 R\rceil}$ CNOT gates \cite{iten2016}.
For qubits ($d=2$), we have $R=k$, so $N_{\rm iso}=\Theta(k\,2^{k})$. This is an improvement by a factor $\Theta(2^{k}/k)$ over the QSD. Here $\Theta(\cdot)$ denotes asymptotic scaling up to constant factors.

Both circuits are essentially sequential cascades of uniformly controlled rotations, so their depth scales like their gate count. Restricting all CNOTs to nearest-neighbor interactions on a linear chain increases the QSD gate count by at most a constant factor~\cite{shende2006}. We expect the same for the isometry, which is built from the same uniformly controlled gates~\cite{iten2016}. Because the layouts used on \textit{ibm\_kingston} embed each register as such a chain, the asymptotic scaling carries over to the hardware. For the small registers used here, the advantage is still modest, about a 45\% reduction in two-qubit gates after transpilation at $k=3$, but it grows rapidly with the number of copies.

\emph{\color{blue}Experimental details on the quantum processor.---}All quantum experiments were run on \texttt{ibm\_kingston}, a 156-qubit Heron r2 processor
with heavy-hex connectivity, accessed via the IBM Quantum cloud platform
\cite{quantumprocessor}. At the time of execution the device had median coherence times
$T_1 \approx 250~\mu$s and $T_2 \approx 130~\mu$s, median single- and two-qubit gate errors of
$ \sim 3.3\times10^{-4}$ and $\sim2.0\times10^{-3}$, and a median readout error of
$\sim 8.7\times10^{-3}$. Benchmarking and validation runs were carried out with the
high-performance simulators provided by Qiskit \cite{javadi2024quantum}.
Circuits were submitted through the \emph{Sampler} interface \cite{javadi2024quantum},
which returns raw bitstring distributions, using $20{,}000$ shots per circuit. Noise was
reduced with two standard techniques: \emph{Matrix-free Measurement Mitigation} (M3) \cite{M3ref}
and \emph{Dynamical Decoupling} (DD) \cite{viola1998dynamical, viola1999dynamical,
duan1999suppressing}. M3 restricts the unfolding problem to the subspace spanned by the
observed bitstrings, which is far smaller than the full Hilbert space and therefore cheap
to invert, while DD suppresses dephasing during idle periods through refocusing pulse
sequences.

\emph{\color{blue}Discussion.---} Correction-free teleportation is particularly useful for retrieving stored quantum operations and for distributed processing when conditional corrections at the receiver are unavailable or inconvenient. Multicopy and port-based teleportation protocols address this requirement using different resources, but both involve collective multiqubit measurements that can be challenging to realize experimentally. Their permutation symmetries provide useful structure, and representation-theoretic methods, including Schur--Weyl techniques \cite{Fei2023efficient, Wills2024efficient} in port-based teleportation, enable efficient implementations. Nevertheless, favorable asymptotic scaling does not necessarily translate into shallow circuits on current quantum processors.

Our work provides a practical route from the measurement description to an implementable circuit
and a proof-of-concept realization of the multicopy teleportation method. By completing the measurement construction to a unitary on an appropriately enlarged space, we map the relevant measurement-basis states to computational-basis states, allowing the outcome to be read out using standard measurements. Combining this construction with QSD or isometry decomposition translates the protocol into elementary gates, with isometry decomposition reducing the circuit overhead in our experiments. Hardware connectivity remains an important consideration: on processors with local connectivity, interactions between nonadjacent qubits can require additional routing gates, increasing circuit depth and exposure to noise. Architectures with greater connectivity could reduce this overhead and facilitate the implementation of both multicopy and port-based teleportation.

The experiments presented here with up to a few input copies provide a starting point for investigating the practical resource requirements of correction-free teleportation and quantum program retrieval. On the theoretical side, circuit constructions that more directly exploit measurement symmetries and hardware connectivity could further reduce gate counts, depth, and ancillary requirements. On the experimental side, improvements in gate fidelity and coherence times, together with larger qubit registers supporting deeper circuits, could extend the accessible copy numbers and enable investigations of port-based teleportation, with its additional entangled ports and collective measurement requirements. These developments could help establish the conditions under which the theoretical advantages of these protocols become experimentally accessible on superconducting quantum hardware.

\emph{\color{blue}Conclusions.---}We have experimentally demonstrated multicopy correction-free teleportation and its application to the single- and two-qubit operations (Fig.~\ref{fig:2}) on superconducting quantum hardware. We showed that additional input copies enhance the success probability, consistent with theoretical predictions, although increasing circuit complexity and hardware noise eventually suppress this advantage with more copies. Our results establish the feasibility of the protocols at small scales and provide a basis for further exploration of probabilistic quantum programming and distributed quantum processing as hardware capabilities improve.

\emph{\color{blue}Acknowledgments.---}This work was supported by the Research Council of Finland project 362573 and the Finnish quantum flagship program. This work is part of the Finnish Centre of Excellence in Quantum Materials (QMAT). We acknowledge the use of IBM Quantum Credits for this work. The views expressed are those of the authors, and do not reflect the official  policy or position of IBM or the IBM Quantum team.

\emph{\color{blue}Data availability.---}The source code, simulation parameters, and the experimental data generated during and/or analyzed during the current study are available in the GitHub repository \cite{MyCode2026}.


\bibliography{refs}
\end{document}